\documentclass[12pt,epsf,amssymb]{article} \textheight =23 truecm
\def\lsim{\raise0.3ex\hbox{$<$\kern-0.75em\raise-1.1ex\hbox{$\sim$}}}
\def\gsim{\raise0.3ex\hbox{$>$\kern-0.75em\raise-1.1ex\hbox{$\sim$}}}
\def\noi{\noindent}  \def\bea{\begin{eqnarray}}
\def\eea{\end{eqnarray}} \def\beq{\begin{equation}}
\def\eeq{\end{equation}} 
\def\beeq{\begin{eqnarray}} \def\eeeq{\end{eqnarray}} \def\R{ {\rm R
\kern -.31cm I \kern .15cm}} \def\C{ {\rm C \kern -.15cm \vrule
width.5pt \kern .12cm}} \def\Z{ {\rm Z \kern -.27cm \angle \kern
.02cm}} \def\N{ {\rm N \kern -.26cm \vrule width.4pt \kern .10cm}}
\def\1{{\rm 1\mskip-4.5mu l} }

\usepackage{graphics} \usepackage{graphicx} \usepackage{epsfig}

\begin{document} 

\begin{center} 

{\Large \bf Leptogenesis and Dirac neutrino masses\par in $SO(10)$}

\author{R. Aleksan}

\author{L. Oliver}

\end{center}

\par \vskip 5 truemm

\begin{center}

 R. Aleksan $^a$ and L. Oliver $^b$

\vskip 2 truemm

$^a$ {\it IRFU, CEA, Universit\' e Paris-Saclay,
91191 Gif-sur-Yvette Cedex, France}

\vskip 2 truemm

$^b$ {\it IJCLab, P\^ole Th\'eorie, CNRS/IN2P3 et Universit\'e Paris-Saclay\par 
b\^at. 210, 91405 Orsay, France}

\vskip 5 truemm

\end{center}

\vskip 8 truemm

\begin{abstract}

\noi Within the scheme of Baryogenesis through Leptogenesis in the single flavor approximation, we emphasize the role of the Dirac neutrino mass $m_D$ in the SO(10) Grand Unification scheme, using the Higgs representations for the fermions masses ${\bf 10}$ and $\overline{\bf 126}$. Both representations are needed to get relations among the fermion masses, e.g. $m_b = m_\tau$, $m_\mu = 3 m_s$,... Taking these two representations allow to relax the condition $m_D = m_u$ obtained with the representation ${\bf 10}$ alone and to have general Dirac masses $m_D \not = m_u$. On the other hand, we assume that the antisymmetric representation ${\bf 120}$ is absent. Within this hypothesis, the unitary matrices $V^L$ and $V^R$ that diagonalize $m_D$ through $m_D = V^{L \dagger} m_D^{diag} V^R$ still satisfy the relation $V^R = V^{L*}$. Under these conditions, unlike the case of keeping only the ${\bf 10}$ representation, we obtain in parameter space several examples of the heavy neutrino spectrum consistent with the SO(10) unification scale, and a value for the baryon asymmetry $Y_B$ in agreement with data. For illustration, we detail a case that gives the heavy spectrum with masses $M_1 = 6.26 \times 10^{10}\ {\rm GeV}$, $M_2 = 2.23 \times 10^{12}\ {\rm GeV}$, $M_3 = 4.57 \times 10^{14}\ {\rm GeV}$ and, in the strong washout regime, the baryon asymmetry $Y_B = 8.50 \times 10^{-11}$ consistent with the data.  

\end{abstract}

\newpage \pagestyle{plain}

\section{Introduction}

This paper is devoted to Baryogenesis via Leptogenesis in the single flavor approximation. In Sections 2 to 4 we provide simple known reviews, that are necessary for clarity, while in  Section 5 we concentrate on the phenomenology of Leptogenesis within SO(10), the main topic of the present work. In Section 4 we review the data on the cosmological baryon-antibaryon asymmetry and the heavy neutrino interactions that lead to a cosmological CP asymmetry and hence to a baryon asymmetry. Section 5 is devoted to the phenomenology within $SO(10)$ inspired models. Among the Higgs representations contributing to the fermion masses, enumerated in Section 3, we begin with details in Section 5.1 of the representation ${\bf 10}$, that implies the equality $m_D = m_u$, and we emphasize the problems found in this case. We then consider in Section 5.2 the two symmetric representations ${\bf 10}$ and $\overline{\bf 126}$ and assume the absence of the antisymmetric representation ${\bf 120}$. This allows to have general Dirac masses $m_D \not = m_u$. We examine in detail the theoretical consistency of this case, and expose its very interesting phenomenological results on heavy neutrino masses and baryon asymmetry.

\section{Neutrino masses and mixing and seesaw model}
\subsection{Light neutrino data and PMNS mixing matrix}
\noi General limits on neutrino masses \cite{PDG}
\beq
\label{1e}
m_{\nu_e} < 0.8\ {\rm eV} \qquad ({\rm from\ nuclear\ \beta\ decay})
\eeq
\beq
\label{2e}
|< m_{ee} >|\ < 0.4\ {\rm eV} \qquad ({\rm from\ neutrinoless \ 2\beta\ decay})
\eeq
\beq
\label{3e}
\sum_i m_{\nu_i} < 0.072\ {\rm eV} \qquad ({\rm from\ astrophysics})
\eeq

\vskip 5 truemm

\noindent The neutrino mass matrix $m_L$ is diagonalized by the $3 \times 3$ Pontecorvo \cite{PONTECORVO}, Maki, Nagakawa and Sakata \cite{MNS} (PMNS) mixing unitary matrix $U$
\beq
\label{4e}
m_L = U \textrm{\ diag}(m_1,m_2,m_3)\ U^\dagger
\eeq

\noi where $U$ is given by 
\beq
\label{5e}
U = \left(
        \begin{array}{ccc}
        c_{12}c_{13} & s_{12}c_{13} & s_{13} e^{-i\delta} \\
        -s_{12}c_{23}-c_{12}s_{23}s_{13} e^{i\delta} & c_{12}c_{23}-s_{12}s_{23}s_{13} e^{i\delta} & s_{23}c_{13} \\
        s_{12}s_{23}-c_{12}c_{23}s_{13} e^{i\delta} & -c_{12}s_{23}-s_{12}c_{23}s_{13} e^{i\delta} & c_{23}c_{13} \\
        \end{array} \right) .\ \textrm{diag}(1,e^{i\alpha},e^{i\beta})
\eeq
\noi $\alpha, \beta$ being Majorana phases.\par

The light neutrinos are generally assumed to be Majorana fields and the phases $\alpha, \beta$ appear for neutrinos because a Majorana mass operator, unlike a Dirac mass operator, is not invariant under a phase redefinition of the fields. Indeed, in the case of a Majorana neutrino field $\nu_R$, the mass operator reads
\beq
\label{6e}
- {1 \over 2} \ \overline{\nu}_R^c m \nu_R + h.c. \ , \qquad \qquad \nu_R^c = C \overline{\nu}_R^T \ , \qquad C = i\gamma^2\gamma^0
\eeq

\noi that by a phase transforms as
\beq
\label{7e}
\nu_R \to e^{i\theta} \nu_R \ , \qquad \qquad \overline{\nu}_R^c m \nu_R \to e^{2i\theta} \ \overline{\nu}_R^c m \nu_R
\eeq

The experimental results from solar (s) and atmospheric (a) neutrinos read \cite{PDG}
\noi 
\beq
\label{8e}
\Delta m_s^2 \simeq 8 \times 10^{-5}\ {\rm eV}^2
\eeq
\beq
\label{9e}
\tan^2 \theta_s \simeq 0.4
\eeq
\beq
\label{10e}
\Delta m_a^2 \simeq 2.5 \times 10^{-3}\ {\rm eV}^2
\eeq
\beq
\label{11e}
\tan^2 \theta_a \simeq 1
\eeq
\beq
\label{12e}
\tan^2 \theta_{13} \simeq 0.02 \qquad (\rm{Daya\ Bay})
\eeq
\noi Taking into account these data on solar and atmosperic neutrinos and the fact that $s_{13}$ is bounded to be small, we will take
\beq
\label{13e}
U \simeq \left(
        \begin{array}{ccc}
       c_s & s_s & 0 \\
        -\ {s_s \over \sqrt{2}} & {c_s \over \sqrt{2}} & {1 \over \sqrt{2}} \\
        {s_s \over \sqrt{2}} & -\ {c_s \over \sqrt{2}} & {1 \over \sqrt{2}} \\
        \end{array}
        \right).\ \textrm{diag}(1,e^{i\alpha},e^{i\beta})
\eeq

\noi i.e.

\beq
\label{14e}
m_L \simeq \left(
        \begin{array}{ccc}
       c_s & s_s & 0 \\
        -\ {s_s \over \sqrt{2}} & {c_s \over \sqrt{2}} & {1 \over \sqrt{2}} \\
        {s_s \over \sqrt{2}} & -\ {c_s \over \sqrt{2}} & {1 \over \sqrt{2}} \\
        \end{array}\right)\ \textrm{diag}(m_1,e^{-2i\alpha}m_2,e^{-2i\beta}m_3)\ \left(
        \begin{array}{ccc}
       c_s & -\ {s_s \over \sqrt{2}} & {s_s \over \sqrt{2}} \\
         s_s & {c_s \over \sqrt{2}} & -\ {c_s \over \sqrt{2}} \\
        0 & {1 \over \sqrt{2}} & {1 \over \sqrt{2}} \\
        \end{array}\right)
\eeq

Notice that, due to the solar and neutrino data, the hierarchy of the angles is quite different than in the quark sector.\par

\subsection{The seesaw model}
\noi Since the mass operator flips helicity, in the case of a Dirac particle one has
\beq
- \overline{\nu}_R m_D \nu_L + h.c.
\label{15e}
\eeq
\noi and in the case of a Majorana neutrino field $N_R$ the mass operator reads
\beq
- {1 \over 2} \ \overline{\nu}_R^c M_R \nu_R + h.c. \ , \qquad \qquad \nu_R^c = C \overline{\nu}_R^T \ , \qquad C = i\gamma^2\gamma^0
\label{16e}
\eeq
\noi The most natural framework to account for the order of magnitude of neutrino masses is the seesaw model \cite{SS}. In this model, the mass part of the Lagragian has the form, for three generations
\beq
\left( \overline{\nu}_L^c \ , \ \overline{\nu}_R^c \right)
\left( \begin{array}{cc}
0 & m_D^t \\
m_D & M_R
\end{array} \right)
\left( \begin{array}{c} 
\nu_L \\ \nu_R 
\end{array} \right) + h.c.
\label{17e}
\eeq
																								
\noi where the $3 \times 3$ Dirac neutrino mass matrix $m_D$ has elements of the order of the masses of charged fermions and $M_R$ is the $3 \times 3$ Majorana mass matrix of the right-handed neutrinos, which are singlets of the Standard Model gauge group, with elements of the order of the scale of breaking of the lepton quantum number.\par
\noi Diagonalizing the $6 \times 6$ seesaw matrix 
\beq
\label{18e}
\left( \begin{array}{cc}
0 & m_D^t \\
m_D & M_R
\end{array} \right)
\eeq
\noi one gets 3 light eigenvalues
\beq
\label{19e}
m_L \simeq -m_D\ M_R^{-1}\ m_D^t
\eeq

\noi and 3 heavy eigenvalues of the order $M_R$,
\beq
\label{20e}
m_H \simeq M_R
\eeq

\noi Notice that the Dirac mass is an arbitrary $3 \times 3$ complex matrix that can be diagonalized by two unitary matrices
\beq
\label{21e}
m_D = V^{L\dagger}\ m_D^{diag}\ V^R
\eeq

\noi Redefining
$$e^{-2i\alpha} m_2 \to m_2\ , \qquad e^{-2i\beta} m_3 \to m_3$$

\noi one can consider the inverse seesaw formula to guess the magnitudes of the $M_R$ masses, 
\beq
\label{22e}
M_R = -m_D^t\ m_L^{-1}\ m_D
\eeq
\noi and inverting the matrix (\ref{19e}) one has
\beq
\label{23e}
m_L^{-1} =  \left(
        \begin{array}{ccc}
        {c_s^2 \over m_1} + {s_s^2 \over m_2} & - {c_s s_s \over \sqrt{2}} \left ({1 \over m_1} - {1 \over m_2} \right ) & {c_s s_s \over \sqrt{2}} \left ({1 \over m_1} - {1 \over m_2} \right ) \\  - {c_s s_s \over \sqrt{2}} \left ({1 \over m_1} - {1 \over m_2} \right )  & {1 \over 2}  \left ({s_s^2 \over m_1} + {c_s^2 \over m_2} + {1 \over m_3} \right ) &  - {1 \over 2}  \left ({s_s^2 \over m_1} + {c_s^2 \over m_2} - {1 \over m_3} \right ) \\   {c_s s_s \over \sqrt{2}} \left ({1 \over m_1} - {1 \over m_2} \right ) &  - {1 \over 2}  \left ({s_s^2 \over m_1} + {c_s^2 \over m_2} - {1 \over m_3} \right ) &  {1 \over 2}  \left ({s_s^2 \over m_1} + {c_s^2 \over m_2} + {1 \over m_3} \right ) \\
        \end{array}
        \right)
 \eeq

\noi To get the magnitude of the heavy right-handed neutrinos one needs therefore the Dirac matrix $m_D$ and the inverse of the light neutrino mass matrix $m_L^{-1}$ and then diagonalize the matrix $M_R$ (\ref{22e}), as we will see below.

\section{Fermion masses in $SO(10)$}

The historically very important grand unified theory that has L-R symmetry and therefore can implement the seesaw mechanism is the $SO(10)$ Grand Unification Model. \par
In this model, L and R fermions of a single generation belong to the same spinorial {\bf 16} representation, e.g. for negative helicity fermions :
\beq
\label{24e} 
\left[ u_L^\alpha, d_L^\alpha, (u_R^\alpha)^c, (d_R^\alpha)^c, \nu_L, e_L, (N_R)^c, (e_R)^c \right] \sim {\bf 16}
\eeq
where $\alpha$ is the color index and the gauge bosons are in the adjoint representation {\bf 45}.\par
Let us now make some remarks on masses and mixing in some particular cases in the interesting $SO(10)$ case. Let us look at the product 
\beq
\label{25e} 
{\bf 16} \times {\bf 16} = {\bf 10}_S + {\bf \overline{126}} _S+ {\bf 120}_A
\eeq

\noindent where ${\bf 10}+{\bf \overline{126}}$ is the symmetric part and {\bf 120} the antisymmetric part. The representatios ${\bf 10}$ and ${\bf 120}$ are real, ${\bf 126}$ is complex, and the Yukawa terms that can give mass to the fermions are
\beq
\label{26e} 
{\bf 16}_f \times {\bf 16}_f \times {\bf 10}_H = {\bf 1} +\ ...
\eeq
\beq
\label{27e} 
{\bf 16}_f \times {\bf 16}_f \times {\bf \overline{126}}_H = {\bf 1} +\ ...
\eeq
\beq
\label{28e} 
{\bf 16}_f \times {\bf 16}_f \times {\bf 120}_H = {\bf 1} +\ ...
\eeq

The Yukawa part of the Lagrangian reads \cite{DILUZIO}
\beq
\label{29e} 
{\cal{L}}_Y = {\bf 16}_f\left(Y_{10}{\bf 10}_H + Y_{126}{\bf \overline{126}}_H + Y_{120}{\bf 120}_H\right){\bf 16}_f
\eeq

\noindent where a possible sum over Higgs representations and Yukawa coupling matrices in family space is implicit. After spontaneous symmetry breaking one gets the masses
$$m_d = v^d_{10}Y_{10} + v^d_{126}Y_{126} + v^d_{120}Y_{120} \ \ $$
$$m_u = v^u_{10}Y_{10} + v^u_{126}Y_{126} + v^u_{120}Y_{120} \ \ $$
\beq
\label{30e} 
m_e = v^d_{10}Y_{10} - 3 v^d_{126}Y_{126} + v^e_{120}Y_{120}
\eeq
$$m_D = v^u_{10}Y_{10} - 3 v^u_{126}Y_{126} + v^D_{120}Y_{120}$$

\noindent where the Yukawa matrices $Y_{10}$ and $Y_{126}$ are complex symmetric, $Y_{120}$ is complex antisymmetric, and the $v$'s are Higgs vacuum expectation values. From the term (\ref{26e}) alone we obtain the well-known relations $m_e = m_d$ and $m_D = m_u$, while the term (\ref{27e}) alone would give the relations $m_e = -3m_d$ and $m_D = -3m_u$, \noindent and {\it no relation} from the term (\ref{28e}).\par
The vev's in (\ref{30e}) are in all generality complex numbers if we assume that $CP$ can be spontaneously broken (soft $CP$ violation). If $CP$ is not spontaneously broken the vevs are real and all $CP$ violation comes from the Yukawa couplings (hard $CP$ violation).\par
One could wonder how within SO(10) one can get the most general counting of parameters done above, i.e. 84(42) parameters for the whole mass sector, with 48(24) parameters in the lepton sector (see \cite{FALCONE-OLIVER}). As said above, this is simply achieved if all the representations ${\bf 10}_H, {\bf \overline{126}}_H, {\bf 120}_H$ in (\ref{29e}) are present and are different for each mass matrix, that becomes then completely general.\par 

An interesting particular case is to consider only the ${\bf 10}$ and ${\bf \overline{126}}$ representations in (\ref{30e}), with ${\bf 120}$ absent :
$$m_d = m^d_{10} + m^d_{126} \ \ $$
$$m_u = m^u_{10} + m^u_{126} \ \ $$
\beq
\label{31e} 
m_e = m^d_{10} - 3 m^d_{126}
\eeq
$$m_D = m^u_{10} - 3 m^u_{126}$$

\noindent In this situation, all fermion mass matrices $m_u, m_d, m_D$ and $m_e$ are complex symmetric.\par

Let us count again the number of parameters under this hypothesis. The complex symmetric matrices $m^d_{10}, m^d_{126}, m^u_{10}, m^u_{126}, m^R_{126}$, have 12(6) parameters each, that gives a total number of 60(30) parameters, a reduction relatively to the 84(42) total number of parameters of the general case. 
One can diagonalize the complex symmetric matrices $m_d,... M_R$ with unitary matrices $V_d,... V_R$. 
Because of the relations (\ref{31e}), the unitary matrices $V_e$, $V_D$, $V_R$ are in principle given in terms of $V_u$ and $V_d$ and mass eigenvalues. Notice that, as discussed in the mass basis for the pure lepton sector, we can adopt without loss of generality the basis in which $m_e = m_e^{diag}$. However, these relations give complicated equations between the elements of mixing matrices. Within this case of considering both ${\bf 10}$ and  ${\bf \overline{126}}$, it seems hard to find relations between the mixing matrices in the quark and the lepton sector, at least in a model-independent way.\par 

Let us consider two limiting cases : only the ${\bf 10}$ or only the ${\bf \overline{126}}$ contribute to $m_d$, $m_u$, $m_e$ and $m_D$.\par

From (\ref{31e}) we see that in both cases one has quark-lepton symmetry in the mixing matrices, i.e. a relation between the left-handed neutrino Dirac mixing matrix $V_L$, where $m_D = V_L^\dagger m_D^{diag} V_R$, and the CKM quark matrix 
\beq
\label{32e} 
V_L = V_uV_d^\dagger = V_{CKM}
\eeq

\noindent This relation has been often used in a number of phenomenological schemes \cite{AKHMEDOV}, \cite{BUCCELLA-FALCONE-OLIVER}. However, as it is well known, one needs both representations ${\bf 10}$ and ${\bf \overline{126}}$ to describe fermion masses in $SO(10)$ \cite{GEORGI-JARLSKOG}\cite{HARVEY-RAMOND-REISS}\cite{BERTOLINI}\cite{JOSHIPURA-PATEL}, and therefore we must conclude that there is a clash between  a good description of fermion masses and the one of obtaining quark-lepton symmetry in mixing.\par
 
\section{Leptogenesis and baryon-antibaryon asymetry}

\subsection{Baryon-antibaryon asymmetry from WMAP}
To get the right relative abundance of the primordial light nuclei one needs the baryon-antibaryon asymmetry (WMAP)
\beq
\label{33e} 
\eta_B = {n_B - n_{\overline{B}} \over n_\gamma} = (6.1 \pm 0.3) \times 10^{-10}
\eeq

\noi The experimental baryon to entropy fraction $Y_B$ is given by WMAP,
\beq
\label{34e} 
Y_B = {n_B-n_{\overline{B}} \over s} \simeq 9 \times 10^{-11}
\eeq

\noi that is smaller than the baryon to photon ratio by roughly a factor 7.\par

The desired baryonic asymmetry may arise from the leptogenesis scenario \cite{LEPTOGENESIS}, with a leptonic asymmetry produced at a high scale, which gives rise by $B-L$ conserving sphaleron processes \cite{KRS} at the electroweak scale to $B$ and $L$ asymmetries below that scale. 

\subsection{Relation between baryon and lepton asymmetries}

An excellent complete and pedagogical account of this important aspect of Baryogenesis via Leptogenesis has been given by Buchm$\tilde {\rm u}$ller, Peccei and Yanagida \cite{PECCEI}, that we summarize now.

In the plasma one assigns a chemical potential $\mu$ to each to the quark, lepton and Higgs fields, i.e. $5N_f+1$ chemical potentials
\beq
\label{35e} 
\mu_{q_i} \ , \qquad \mu_{u_i} \ , \qquad \mu_{d_i} \ , \qquad \mu_{\ell_i} \ , \qquad \mu_{e_i} \qquad (i = 1,... N_f) \ , \qquad \mu_H
\eeq

\noi The asymmetry in the particle and antiparticle number densities read
\beq
\label{36e} 
n_i - \overline{n}_i = {gT^3 \over 6} \beta\mu_i \qquad ({\rm fermions}) \ , \qquad n_i - \overline{n}_i = {gT^3 \over 6} 2\beta\mu_i \qquad ({\rm bosons})
\eeq

\noi The constraints on chemical potentials in thermal equilibrium for $100\ {\rm GeV} < T < 10^{12}\ {\rm GeV}$ are obtained from the following theoretical conditions.\par 
The crucial $SU(2)$ electroweak instantons (sphalerons) relate lepton and quark chemical potentials
\beq
\label{37e} 
\sum_i (3\mu_{q_i} + \mu_{\ell_i}) = 0
\eeq

\noi On the other hand, adding the relations obtained from QCD instantons, from the gauge and Yukawa interactions, and imposing the vanishing of the total hypercharge, one obtains, from the expressions of baryon and lepton number densities $n_B = gBT^2/6, n_L = g L T^2/6$,
\beq
\label{38e} 
B = \sum_i (2\mu_{q_i} + \mu_{u_i} + \mu_{d_i}) \ , \qquad  L = \sum_i (2\mu_{\ell_i} + \mu_{e_i})
\eeq

\noi and one finds, 
\beq
\label{39e} 
B = c_s (B-L) \ , \qquad  L = (c_s-1)(B-L)
\eeq

\noi with
\beq
\label{40e} 
c_s = {8N_f + 4 \over 22 N_f + 13}
\eeq

\noi i.e., within the leptogenesis scenario, the baryon to entropy fraction is given by
\beq
\label{41e} 
B \simeq - {1 \over 2} L
\eeq

\noi Relations (\ref{39e}) suggests that (B-L)-violation is needed to generate a B-asymmetry.

\subsection{Heavy neutrino interactions} 
\noi The decay of the lightest heavy neutrino $N_1$ violating CP and leading to leptogenesis proceeds through tree and loop diagrams involving the Standard Model Higgs field, that will give the Dirac mass $m_D$.\par

\noi Fermion-Higgs Yukawa interactions, with sum over families $i, j = 1, 2, 3$ are given by
\beq
\label{42e} 
L_H = \sum_{i,j} \left( f_{ij} \overline{e}_{Ri} \ell_{Lj} H^\dagger + h_{ij} \overline{N}_Ri \ell_{Lj} H \right)
\eeq

\noi The decay width of the heavy neutrino $N_i$ at tree level is given by
\beq
\label{43e} 
\Gamma_{Di} = \Gamma(N_i \to H \ell_L) + \Gamma(N_i \to H^\dagger \ell_L^\dagger) = {1 \over 8 \pi} (h^\dagger h)_{ii} M_i
\eeq

\noi The complex {\it symmetric} right-handed neutrino mass matrix is diagonalized by a single unitary matrix $W_R$,
\beq
\label{44e} 
M_R^{diag} = - W_R^\dagger m_D^t m_L^{-1} m_D W_R^* = - \hat{m}_D^t m_L^{-1} \hat{m}_D
\eeq

\noi Denoting by $M_1$ the mass of the lightest heavy neutrino, one gets the CP asymmetry parameter $\epsilon_1$ 
\beq
\label{45e} 
\epsilon_1 = {1 \over {4 \pi}} \sum_{k \not =1} f\left({M_k^2 \over M_1^2} \right) {{\rm Im}\left\{\left[(h^\dagger h)_{1k}\right]^2 \right \} \over (h^\dagger h)_{11}}
\eeq
\beq
\label{46e} 
f(x) = \sqrt{x}\left[ {1 \over 1-x} + 1 - (1+x) {\rm ln} \left ({1+x \over x}\right) \right] \qquad \to \qquad -{3 \over {2\sqrt{x}}} \ \ \ \ (x >> 1)
\eeq

\noi In terms of the Dirac mass in the basis of diagonal $M_R$ (\ref{44e})
\beq
\label{47e} 
\hat{m}_D = m_D W_R^*
\eeq

\noi the CP asymmetry (\ref{45e}) writes, for the lightest heavy neutrino $N_{R_1}$, from (\ref{45e})
\beq
\label{48e} 
\epsilon_1 = {1 \over {4 \pi v^2}} \sum_{k \not =1} f\left({M_k^2 \over M_1^2} \right) {{\rm Im}\left \{ \left[(\hat{m}_D^\dagger\hat{m}_D)_{1k}\right]^2 \right \} \over (\hat{m}_D^\dagger\hat{m}_D)_{11}}
\eeq

\noi In the preceding formulas $m_D = h v$, with $v =\ <H>$.

\subsection{CP asymmetry leads to (B-L)-asymmetry}

\noi The CP-asymmetry leads to a L-asymmetry or, equivalently, to a (B-L)-asymmetry
\beq
\label{49e} 
Y_{B-L} \simeq - Y_L = - {n_L-n_{\overline{L}} \over s} = - \kappa {\epsilon_1 \over g_*}
\eeq

\noi where $s$ is the entropy $s = 7.04\ n_\gamma$, and $g_* \sim 100$ is the number of degrees of freedom in the plasma. The efficiency factor $\kappa < 1$ takes into account the washout processes. To get $\kappa$ one needs to solve the Boltzmann equations.\par

\subsection{Departure from thermal equilibrium}

\noi Leptogenesis takes place at temperatures $T \sim M_1$.\par

\noi Comparing the decay width $\Gamma_1(T)$ with the Hubble parameter $H(T)$ we have\par 
\noi - for $\Gamma_1(T) < H(T)$, heavy neutrinos are out of thermal equilibrium,\par
\noi - for $\Gamma_1(T) > H(T)$, heavy neutrinos are in thermal equilibrium,\par
\noi and the effective neutrino mass, that controls the amount of washout is given by
\beq
\label{50e} 
{\tilde m}_1 = {(\hat{m}_D^\dagger\hat{m}_D)_{11} \over M_1}
\eeq

\noi The borderline between the two regimes, using $H(T] = 1.66 g_* T^2 /M_P$, $g_* = g_{SM} = 106.75$, is given by 
\beq
\label{51e} 
\Gamma_1 = H\mid_{T = M_1} \qquad \to \qquad {\tilde m}_1 = m_*
\eeq

\noi with
\beq
\label{52e} 
m_* = {16 \pi^{5/2} \over 3\sqrt{5}} g_*^{1/2} {v_F^2 \over M_P} \simeq 10^{-3}\ {\rm eV}
\eeq

\noi It is quite remarquable that the equilibrium neutrino mass $m_*$ is close to the neutrino masses suggested by neutrino oscillations
\beq
\label{53e} 
\sqrt{\Delta m_{sol}^2} \simeq 8 \times 10^{-3}\ {\rm eV} \ , \qquad \sqrt{\Delta m_{atm}^2} \simeq 5 \times 10^{-2}\ {\rm eV} 
\eeq

\noi The Boltzmann equations for Leptogenesis read \cite{BUCHMULLER-BARI-PLUMACHER}
\beq
\label{54e} 
{dN_{N_1} \over dz} = -(D+S)(N_{N_1}-N_{N_1}^{eq})
\eeq
\beq
\label{55e} 
{dN_{B-L} \over dz} = - \epsilon_1 D (N_{N_1}-N_{N_1}^{eq}) - W N_{B-L}
\eeq

\noi where $z = M_1/T$, the number density is $N_{N_1}$, $N_{N_1}^{eq} (z<<1) = 3/4$ and the amount of $B-L$ asymmetry is denoted by $N_{B-L}$. In the preceding equations, $D$ means decay, $S$ means scattering and $W$ means washout.\par
The evolution of $N_1$ abundance and the $B-L$ asymmetry gives a typical choice of parameters, $M_1 = 10^{10}\ {\rm GeV}$, $\epsilon_1 = 10^{-6}$, ${\tilde m}_1 = 10^{-3}\ {\rm eV}$ and $\overline{m} = \sqrt{m_1^2+m_2^2+m_3^2} = 0.05\ {\rm eV}$.

Keeping only the decay $D$ is a good approximation of the general case. The parameter
\beq
\label{56e} 
K = {\Gamma_D(z = \infty) \over H(z = 1)} = {{\tilde m}_1 \over m_*}
\eeq
controls the amount of washout, weak washout for $K << 1$ (regime far out of equilibrium) and strong washout for $K >> 1$ (neutrino abundance close to equilibrium).\par
	Both the scales of solar and atmospheric neutrino oscillations $\sqrt{\Delta m_{sol}^2} \simeq 8 \times 10^{-3}\ {\rm eV}$, $\sqrt{\Delta m_{atm}^2} \simeq 5 \times 10^{-2}\ {\rm eV}$ are larger than the equilibrium neutrino mass $m_*$. The range of neutrino masses and therefore ${\tilde m}_1$ indicated by neutrino oscillations lies entirely in the strong washout regime where theoretical uncertainties are small and the efficiency factor $\kappa_f $ is still large enough to allow for successful Leptogenesis.

\noi In the case of strong washout
\beq
\label{57e} 
{\tilde m}_1 >> 3 \times 10^{-3}\ {\rm eV}
\eeq

\noi one gets the baryon asymmetry, {\it in the one-flavor approximation},
\beq
\label{58e} 
Y_{B_1} = - {1 \over 2} \ 0.3 \ {\epsilon_1 \over g_*} \left({0.55 \times 10^{-3} {\rm eV} \over {\tilde m}_1} \right)^{1.16}
\eeq

\noi where $g_* = 107$ in the Standard Model.

\section{Phenomenology within $SO(10)$ inspired models}

\subsection{Phenomenology in the case of Higgs representation ${\bf 10}$ for fermion masses that implies $m_D = m_u$, $V^R = V^{L*}$ and $V^L = V_{CKM}$}
 
\noi Keeping only the ${\bf 10}$ Higgs representation for fermion masses we adopt the diagonalized Dirac neutrino mass matrix
\beq
\label{59e}
m_D^{diag} = \left(
        \begin{array}{ccc}
       m_{D_1} & 0 & 0 \\
        0 & m_{D_2} & 0 \\
        0 & 0 & m_{D_3} \\
        \end{array}
        \right)
\eeq
and the numerical values proposed in \cite{AKHMEDOV}, inspired from the up-quark mass matrix
\beq
\label{60e}
m_{D_1} = 10^{-3}\ {\rm GeV} \qquad \qquad m_{D_2} = 0.4\ {\rm GeV} \qquad \qquad m_{D_3} = 100\ {\rm GeV}
\eeq

Within $SO(10)$, with the electroweak Higgs boson belonging to the {\bf 10} and/or {\bf 126} representations, and {\it no component along the {\bf 120} representation}, the mass matrices are symmetric. As a consequence, the unitary matrices $V^R$ and $V^L$ that diagonalize the Dirac neutrino matrix (\ref{21e}) are related by $V^R = V^{L*}$ and the matrix $M_R$ (\ref{22e}) reads
\beq
M_R = -m_D^t\ m_L^{-1}\ m_D = -\ V^{L+}\ m_D^{diag}\ V^{L*}\ m_L^{-1}\ V^{L+}\ m_D^{diag}\ V^{L*}
\label{61e}
\eeq

For $V^L$ we will assume a form qualitatively similar to the Cabibbo-Kobayashi-Maskawa (CKM) quark matrix, a correct hypothesis at least for exact quark-lepton symmetry, with Higgs in the {\bf 10} representation,
\beq
\label{62e}
V^L = V_{CKM}
\eeq

\noi Of course, in our problem the Wolfenstein parameters $\lambda, A, \rho, \eta$ do not necessarily have the same precise values as in the quark sector : we are interested only in an order of magnitude estimate.\par

\noi Using the parameters 
\beq
\label{63e} 
\lambda = 0.14 \ , \qquad A = 0.8 \ , \qquad \rho = 0.13 \ , \qquad \eta = 0.35
\eeq

\noi and from
\beq
\label{64e} 
s_{12} = \lambda \ , \qquad s_{23} = A \lambda^2 \ , \qquad s_{13} e^{i \delta_L} = A \lambda^3 (\rho+i\eta)
\eeq

\noi one has the mixing matrix 
\beq
\label{65e} 
V^L \simeq \left(
        \begin{array}{ccc}
       0.990 & 0.140 & (2.8-7.7i) \times 10^{-4} \\
        -0.140 & 0.990 & 0.016 \\
        (19.1-7.7i) \times 10^{-4} & -0.016 & 1. \\
        \end{array}
        \right)
\eeq

\noi For the light neutrino spectrum, neglecting small Majorana phases, we adopt the values
\beq
\label{66e} 
m_1 = 0.0030\ {\rm eV} \ , \qquad \qquad m_2 = 0.0095\ {\rm eV} \ , \qquad \qquad m_3 = 0.0495\ {\rm eV}
\eeq

\noi that gives values consistent with the solar and atmospheric data (\ref{8e}),(\ref{10e}).
\noi We adopt the mixing angle
\beq
\label{67e} 
{\rm tan}\theta _s = \sqrt{0.4}
\eeq

\noi that, with (\ref{13e}), gives the mixing matrix
\beq
\label{68e}
U \simeq \left(
        \begin{array}{ccc}
       0.85 & 0.53 & 0 \\
        - 0.38 & 0.60 & 0.71 \\
        0.38 & - 0.60 & 0.71 \\
        \end{array}
        \right)
\eeq
\noi The neutrino mass matrix is then given by 
\beq
\label{69e}
m_L = U^* m_L^{diag} U^\dagger = \left(
        \begin{array}{ccc}
       4.86 \times 10^{-12} & 2.08 \times 10^{-12} & -2.08 \times 10^{-12} \\
        2.08 \times 10^{-12} & 2.86 \times 10^{-11} & 2.09 \times 10^{-11} \\
        - 2.08 \times 10^{-12} & 2.09 \times 10^{-11} & 2.86 \times 10^{-11} \\
        \end{array}
        \right) \ {\rm GeV}
\eeq
\noi and its inverse by
\beq
\label{70e}
m_L^{-1} = \left(
        \begin{array}{ccc}
       2.68 \times 10^{11} & - 7.29 \times 10^{10} & 7.29 \times 10^{10} \\
        - 7.29 \times 10^{10} & 9.53 \times 10^{10} & - 7.51 \times 10^{10} \\
        7.29 \times 10^{10} & - 7.51 \times 10^{10} & 9.53 \times 10^{10} \\
        \end{array}
        \right) \ {\rm GeV^{-1}}
\eeq
\noi Let us now give the complex {\it symmetric} Dirac neutrino mass. We get
$$m_D = V^{L\dagger}\ m_D^{diag}\ V^R = V^{L\dagger}\ m_D^{diag}\ V^{L*}$$
\beq
\label{71e}
= \left(
        \begin{array}{ccc}
        0.0091+0.0003i & -0.0583-0.0012i & 0.1901+0.0768i \\
        -0.0583-0.0012i & 0.4168 & -1.5618 \\
        0.1901+0.0768i & -1.5618 & 100. \\
        \end{array}
        \right)\ {\rm GeV}
\eeq

\vskip 5 truemm

The heavy Majorana neutrino matrix $M_R$ (\ref{22e}) is a complex symmetric matrix and to diagonalize it we need just the single unitary matrix $W_R$
\beq
\label{72e}
M_R = W_R M_R^{diag} W_R^t
\eeq

\noi We find the eigenvalues
\beq
\label{73e}
M_R^{diag} \simeq \left(
        \begin{array}{ccc}
        9.80 \times 10^{14} & 0 & 0 \\
        0 & 6.83 \times 10^9 & 0 \\
        0 & 0 & 1.70 \times 10^5 \\
        \end{array}
        \right)\ {\rm GeV}
\eeq

\noi and the unitary matrix $W_R$
\beq
\label{74e}
W_R \simeq \left(
        \begin{array}{ccc}
        0.0008-0.0024i & -0.0001-0.1422i & -0.0001+0.9898i \\
        0.0191i & 0.0007+0.9897i & 0.1422i \\
        0.0006-0.9998i & 0.0001+0.0193i & -0.0008+0.0003i \\
        \end{array}
        \right)
\eeq

\vskip 3 truemm

In conclusion, labelling 1 and 3 the lighter and heavier neutrino, one gets the rather splitted right-handed heavy neutrino spectrum
\beq
\label{75e}
M_1 = 1.70 \times 10^5\ {\rm GeV} \qquad  M_2 = 6.83 \times 10^9\ {\rm GeV} \qquad M_3 = 9.80 \times 10^{14}\ {\rm GeV} 
\eeq
\noi that writes, in terms of the Dirac mass and the light neutrino mass matrices,
\beq
\label{76e}
M_R^{diag} = - W_R^\dagger\ m_D^t\ m_L^{-1}\ m_D\ W_R^* = - \hat{m}_D^t\ m_L^{-1}\ \hat{m}_D
\eeq

\noi where
\beq
\label{77e}
\hat{m}_D = m_D\ W_R^* \simeq \left(
        \begin{array}{ccc}
        -0.0767 + 0.1913i & 0.0002+0.0553i & -0.0009i \\
        -0.0010-1.5700i & 0.0003-0.3907i & -0.0010i \\
        0.0584+100.0i & 0.0002-0.3530i & -0.0001i \\
        \end{array}
        \right) \ {\rm GeV}
\eeq 

\noi is the Dirac mass matrix (\ref{47e}) in the basis in which the heavy right-handed neutrino mass matrix is diagonal.\par
Taking into account the ordering and notation of the heavy neutrino masses (\ref{75e}), the CP asymmetry (\ref{48e}) will write, for the lightest heavy neutrino
\beq
\label{78e} 
\epsilon_1 = {1 \over {4 \pi v^2}} \sum_{k \not =1} f\left({M_k^2 \over M_1^2} \right) {{\rm Im}\left \{ \left[(\hat{m}_D^\dagger\hat{m}_D)_{1k}\right]^2 \right \} \over (\hat{m}_D^\dagger\hat{m}_D)_{11}}
\eeq

\noi and taking the limit (\ref{46e}) one gets
\beq
\label{79e} 
f\left({M_2^2 \over M_1^2}\right) \simeq - {3 \over {2 \left( {M_2 \over M_1} \right)} } \simeq -3.72 \times 10^{-5} \ , \ \  f\left({M_3^2 \over M_1^2}\right) \simeq - {3 \over {2 \left( {M_3 \over M_1} \right)} } \simeq -2.59 \times 10^{-10}
\eeq

\noi From the matrix (\ref{77e}) one gets finally the CP asymmetry
\beq
\label{80e} 
\epsilon_1 \simeq - 1.32 \times 10^{-13}
\eeq

\noi and from (\ref{50e}), we have
\beq
\label{81e}
{\tilde m}_1 = {(\hat{m}_D^\dagger\hat{m}_D)_{11} \over M_1} \simeq 10^{-2} {\rm eV} > 3 \times 10^{-3} {\rm eV}
\eeq

\noi and therefore we are within the strong washout regime, that gives 
\beq
\label{82e}
Y_{B_1} = - {1 \over 2} \ 0.3 \ {\epsilon_3 \over g_*} \left({0.55 \times 10^{-3} {\rm eV} \over {\tilde m}_3} \right)^{1.16} \simeq 6.39 \times 10^{-18}
\eeq

\noi where $g_* = 107$ in the Standard Model.\par
Therefore, one gets a much too small baryon asymmetry compared with the experimental value (\ref{34e}). The reason is that the leptonic CP asymmetry turns out to be very small for a very splitted heavy neutrino spectrum, that yields to the small value (\ref{80e}).\par
In other terms, this scheme violates the Davidson-Ibarra lower bound on the mass of the lightest heavy right-handed neutrino \cite{DI}.\par
Of course, these numerical results yield for most of the parameter space of the case of the ${\bf 10}$ reprensentation. Akhmedov et al. \cite{AKHMEDOV} go further and look for level crossing points that allow, due to the quasi-degeneracy of the heavy right-handed neutrinos, to have values that are very much larger than (\ref{82e}). However, this feature happens only for very small regions of parameter space. Also, Buccella et al. \cite{BUCCELLA-FALCONE-OLIVER} have proposed a variant of the model with the ${\bf 10}$ representation where the three heavy neutrinos are quasi-degenerate and the model satisfies the Davidson-Ibarra bound. One obtains also in this case a baryon asymmetry consistent with the data. 

\subsection{Case of Higgs representations ${\bf 10}$ and ${\bf \overline{126}}$ for fermion masses implying $m_D \not = m_u$, $V^R = V^{L*}$ and $V^L \not = V_{CKM}$}

We assume now that the fermion masses come from a combination of Higgs bosons belonging to the {\bf 10} and $\overline{{\bf 126}}$ representations that implies the fermion masses
$$m_d = m_{10}^d + m_{126}^d$$
$$m_u = m_{10}^u + m_{126}^u$$
\beq
\label{83e}
m_e = m_{10}^d - 3m_{126}^d
\eeq
$$m_D = m_{10}^u - 3m_{126}^u$$
\noi i.e. at the unification scale one gets an inequality among mass matrices
\beq
\label{84e}
m_D = m_{10}^u + m_{126}^u - 4m_{126}^u = m_u - 4m_{126}^u \not = m_u
\eeq
Therefore, a main consequence is that one cannot assimilate the Dirac neutrino mass matrix $m_D$ to the up-quark mass matrix $m_u$.\par
Let us now turn to the mixing. As pointed out above, in the preceding subsection, within $SO(10)$, with the Higgs bosons belonging to the ${\bf 10}$ and/or ${\bf \overline{126}}$ representations, and {\it no component along the ${\bf 120}$ representation}, the mass matrices are symmetric. As a consequence, the unitary matrices $V^R$ and $V^L$ that diagonalize the Dirac neutrino matrix (\ref{21e}) are related,
\beq
V^R = V^{L*}
\label{85e}
\eeq
and the matrix $M_R$ (\ref{22e}) reads
\beq
M_R = -m_D^t\ m_L^{-1}\ m_D = -\ V^{L+}\ m_D^{diag}\ V^{L*}\ m_L^{-1}\ V^{L+}\ m_D^{diag}\ V^{L*}
\label{86e}
\eeq

From (\ref{83e}), if one keeps \underline{only} the ${\bf 10}$ representation \underline{or only} the ${\bf \overline{126}}$ representation, the left-handed neutrino Dirac mixing matrix $V^L$, where $m_D = V^{L\dagger} m_D^{diag} V^R$ is related to the left-handed up and down quarks matrices from
\beq
\label{87e}
V^L = V^{L}_u V^{L\dagger}_d = V_{CKM}
\eeq

\noi However, as it is well known, and we have pointed out above, one needs both representations ${\bf 10}$ and ${\bf \overline{126}}$ to describe fermion masses in $SO(10)$
\beq
\label{88e}
m_b = m_\tau \ , \qquad \qquad m_\mu = 3 m_s \ ,... \qquad \qquad 
\eeq

Therefore we must conclude that there is a clash between  a good description of fermion masses and the one of obtaining quark-lepton symmetry in mixing, i.e. relation (\ref{88e}), $V^L = V_{CKM}$.\par
In conclusion, if $m_D \not = m_u$ as in (\ref {84e}), this implies also that 
\beq
\label{89e}
m_D \not = m_u \qquad \qquad \to \qquad \qquad V^L \not = V^{L}_u V^{L\dagger}_d = V_{CKM}
\eeq

\noi One must emphasize however that with the Higgs bosons belonging to the ${\bf 10}$ and/or ${\bf \overline{126}}$ representations, the relation $V^R = V^{L*}$, still holds.\par
We now propose a numerical model satisfying (\ref{89e}). From trial and error, among the number of possible models that are successful, we propose the following one.\par

\noi We adopt the diagonalized Dirac neutrino mass matrix
\beq
\label{90e}
m_D^{diag} = \left(
        \begin{array}{ccc}
       m_{D_1} & 0 & 0 \\
        0 & m_{D_2} & 0 \\
        0 & 0 & m_{D_3} \\
        \end{array}
        \right)
\eeq
\noi with the numerical values 
\beq
\label{91e}
m_{D_1} = 0.6\ {\rm GeV} \qquad \qquad m_{D_2} = 5.\ {\rm GeV} \qquad \qquad m_{D_3} = 100.\ {\rm GeV}
\eeq

\noi This spectrum is still hierarchical like the one of proposed in (\ref{60e}), inspired from the up-quark mass matrix, but the values of the light Dirac mass eigenvalues are rather different.

We emphasize again now an important point. Within $SO(10)$, with the electroweak Higgs boson belonging to the {\bf 10} and/or {\bf 126} representations, and {\it no component along the {\bf 120} representation}, the mass matrices are symmetric. As a consequence, the unitary matrices $V^R$ and $V^L$ that diagonalize the Dirac neutrino matrix (\ref{21e}) are related by $V^R = V^{L*}$ and the matrix $M_R$ (\ref{22e}) reads like (\ref{61e}).

For $V^L$ we adopt the following form. 
\beq
\label{92e}
V^L = \left(
        \begin{array}{ccc}
        C_{12}C_{13} & S_{12}C_{13} & S_{13} e^{-i\Delta} \\
        -S_{12}c_{23}-C_{12}s_{23}S_{13} e^{i\Delta} & C_{12}C_{23}-S_{12}S_{23}S_{13} e^{i\Delta} & S_{23}C_{13} \\
        S_{12}s_{23}-C_{12}C_{23}S_{13} e^{i\Delta} & -C_{12}S_{23}-S_{12}C_{23}S_{13} e^{i\Delta} & C_{23}C_{13} \\
        \end{array} \right)
\eeq
$$C_{ij} = \cos(\Theta_{ij})\ , \qquad \qquad S_{ij} = \sin(\Theta_{ij})$$

\noi with
\beq
\label{93e}
\Theta_{12} = 3.140 \ , \qquad \Theta_{23} = 0.500 \ , \qquad \Theta_{31} = 0.400 \ , \qquad \Delta = 1.047
\eeq

\noi that gives
\beq
\label{94e}
V^L \simeq \left(
        \begin{array}{ccc}
       -0.9211 & 0.0015 & 0.1948-0.3372 i \\
        0.0920+0.1617 i & -0.8777-0.0003 i& 0.4416 \\
       0.1717+0.2960 i & 0.479-0.0005 i & 0.8083 \\
        \end{array}
        \right)
\eeq

\noi This matrix is quite different from the CKM matrix, and this is not unexpected due to the remarks concerning (\ref{89e}), that concern the superposition of the representations {\bf 10} and {\bf 126}.

\noi Let us now give the complex {\it symmetric} Dirac neutrino mass. We get
$$m_D = V^{L\dagger}\ m_D^{diag}\ V^R = V^{L\dagger}\ m_D^{diag}\ V^{L*}$$
\beq
\label{95e}
= \left(
        \begin{array}{ccc}
        -5.389-10.3106 i & 7.8365-13.4617 i & 13.9737-24.4633i \\
        7.8365-13.4617 i & 26.8108+0.0429 i & 36.7925+0.0390 i \\
        13.9737-24.4633i & 36.7925+0.0390 i & 66.2655+0.0788 i \\
        \end{array}
        \right)\ {\rm GeV}
\eeq

\vskip 5 truemm

\noi As in the precedent section let us diagonalize the complex symmetric matrix $M_R$
\beq
\label{96e}
M_R = W_R M_R^{diag} W_R^t
\eeq

\noi We find the eigenvalues
\beq
\label{97e}
M_R^{diag} \simeq \left(
        \begin{array}{ccc}
        4.57 \times 10^{14} & 0 & 0 \\
        0 & 2.23 \times 10^{12} & 0 \\
        0 & 0 & 6.26 \times 10^{10} \\
        \end{array}
        \right)\ {\rm GeV}
\eeq

\noi and the unitary matrix
\beq
\label{98e}
W_R \simeq \left(
        \begin{array}{ccc}
        -0.3591+0.0030i & -0.1113-0.0766i & 0.4784-0.7899i \\
       -0.2430-0.3430i & -0.0215+0.9066i & -0.0306+0.0057i \\
        -0.4490-0.7019i & 0.0300-0.3982i & -0.3824+0.0042i \\
        \end{array}
        \right)
\eeq

\vskip 3 truemm

In conclusion, labelling 1 and 3 respectively the lighter and heavier neutrino, one gets the right-handed heavy neutrino spectrum
\beq
\label{99e}
M_1 = 6.26 \times 10^{10}\ {\rm GeV} \qquad M_2 = 2.23 \times 10^{12}\ {\rm GeV}\qquad M_3 = 4.57 \times 10^{14}\ {\rm GeV}
\eeq
\noi that is consistent with the SO(10) unification scale and satisfies the Davidson-Ibarra bound \cite{DI}. This spectrum writes, in terms of the Dirac mass and the light neutrino mass matrices,
\beq
\label{100e}
M_R^{diag} = - W_R^\dagger\ m_D^t\ m_L^{-1}\ m_D\ W_R^* = - \hat{m}_D^t\ m_L^{-1}\ \hat{m}_D
\eeq

\noi where
$$\hat{m}_D = m_D\ W_R^*$$
\beq
\label{101e}
\simeq \left(
        \begin{array}{ccc}
        15.5155 + 30.4706i & -0.8234-1.2499ii & -0.1979+0.4735i \\
        -25.9309+39.8060i & 0.7095-7.5575i & -0.5072-0.5759i \\
        -43.8532+67.8338i & 1.5187-3.1757i & -0.4571-1.1886i \\
        \end{array}
        \right) \ {\rm GeV}
\eeq 

\noi is the Dirac mass matrix (\ref{47e}) in the basis in which the heavy right-handed neutrino mass matrix is diagonal.\par
The CP asymmetry (\ref{45e}) will write, for the lightest heavy neutrino as
\beq
\label{102e} 
\epsilon_1 = {1 \over {4 \pi v^2}} \sum_{k \not =1} f\left({M_k^2 \over M_1^2} \right) {{\rm Im}\left \{ \left[(\hat{m}_D^\dagger\hat{m}_D)_{1k}\right]^2 \right \} \over (\hat{m}_D^\dagger\hat{m}_D)_{11}}
\eeq

\noi and taking the limit (\ref{46e}) one gets
\beq
\label{103e} 
f\left({M_2^2 \over M_1^2}\right) \simeq - {3 \over {2 \left( {M_2 \over M_1} \right)} } \simeq -4.213 \times 10^{-2} \ , \ \  f\left({M_3^2 \over M_1^2}\right) \simeq - {3 \over {2 \left( {M_3 \over M_1} \right)} } \simeq -2.057 \times 10^{-4}
\eeq

\noi From the matrix (\ref{101e}) one gets finally the CP asymmetry
\beq
\label{104e} 
\epsilon_1 \simeq - 8.62 \times 10^{-6}
\eeq

\noi and from (\ref{50e}), we have
\beq
\label{105e}
{\tilde m}_1 = {(\hat{m}_D^\dagger\hat{m}_D)_{11} \over M_1} \simeq 3.95 \times 10^{-2}\ {\rm eV} > 3 \times 10^{-3}\ {\rm eV}
\eeq

\noi and therefore we are within the strong washout regime, that gives 
\beq
\label{106e}
Y_{B_1} = - {1 \over 2} \ 0.3 \ {\epsilon_3 \over g_*} \left({0.55 \times 10^{-3} {\rm eV} \over {\tilde m}_3} \right)^{1.16} \simeq 8.50 \times 10^{-11}
\eeq

\noi where $g_* = 107$ in the Standard Model.\par
Therefore, one gets a value for the baryon asymmetry that is in very good agreement with the experimental value (\ref{34e}).\par
Let us make some remarks on our procedure to obtain the heavy neutrino spectrum (\ref{99e}) and the baryon asymmetry (\ref{106e}).\par Since it is not practical to make a global fit because of the very large number of parameters, we have proceeded in the following way.\par
1) We have first arbitrarily fixed the angle $\Theta_{13} = 0.4$ and the phase $\Delta = 1.047$ of $V^L$, and also $m_{D_1} = 1.5$, $m_{D_2} = 2.5$, $m_{D_3} = 100.$ {\rm GeV}. \par
2) We have scanned the values of $\Theta_{12}$ and $\Theta_{23}$ that give the largest value of $Y_B$, and this gives $\Theta_{12} = 3.14$ and $\Theta_{23} = 0.6$. \par
3) We have varied $\Theta_{23} = 0.6$ to see whether $Y_B$ is very sensitive to this parameter. We have found that this is not the case and we have arbitrarily fixed $\Theta_{23} = 0.5$.\par
4) We have proceeded in the same way for $\Theta_{12}$. We have found that $Y_B$ is moderately sensitive unless this parameter is very close to $\pi$. We have arbitrarily fixed $\Theta_{12} = 3.14$.\par
5) Finally we have scanned $m_{D_1}$ between 0.003 $\rm{GeV}$ and 2. $\rm{GeV}$ and we have looked for the value that gives the experimental value of $Y_B$. We have found $m_{D_1} = 0.6\ \rm{GeV}$.\par
6) We have checked the sensitivity of the final result to the values of $m_{D_2}$ and $m_{D_3}$ and we have found that it is not very sensitive.\par
7) Moreover, if one changes the value of $\Theta_{12}$, for example $\Theta_{12} = 3.$ one finds that one needs $m_{D_1} = 0.9 \ \rm{GeV}$ to get the same value of $Y_B$ and if one chooses $\Theta_{12} = 2.5$ then one needs $m_{D_1} = 3.5 \ \rm{GeV}$ to get the same value of $Y_B$. Therefore $\Theta_{12}$ and $m_{D_1}$ are strongly correlated.

\section{Conclusion}

 In Section 5.1 we have discussed the model based on Dirac masses $m_D = m_u$ and $V^L \simeq V_{CKM}$, suggested by the Higgs representation for fermion masses ${\bf 10}$ within $SO(10)$, in the single flavor approximation. We have concluded that in most of the parameter space the baryon asymmetry turns out to be extremely small. \par
Then, we have proposed a model for Dirac masses in Section 5.2 where the condition $m_D = m_u$ is relaxed, and we have pointed out that this is natural  if one assumes a scheme with the representations ${\bf 10}$ + ${\bf \overline{126}}$. Since both representations are symmetric one keeps still the condicion $V^R = V^{L*}$, where these matrices diagonalize the Dirac neutrino mass $m_D = V^{L\dagger}m_D^{diag}V^R$. However, in this case with two representations, the matrix $V^L$ turns out to be different from $V_{CKM}$.\par
Consistent with this scheme of ${\bf 10}$ + ${\bf \overline{126}}$ Higgs representations for the fermions masses, by trial and error we have looked for a hierarchical spectrum of Dirac masses $m_{D_1}, m_{D_2}, m_{D_3}$ different from the $m_u$ quark spectrum, and a unitary mixing matrix $V^L \not = V_{CKM}$ that give a reasonable heavy right-handed neutrino spectrum consistent with the $SO(10)$ unification scale, with the Davidson-Ibarra bound and a cosmological baryon asymmetry $Y_B$ in agreement with the experimental observations. We have found several solutions for the $m_D$ spectrum and $V^L$, among them one that gives a very reasonable splitted heavy neutrino spectrum at an acceptable scale.\par
We have performed this work in the single flavor approximation and we have left aside flavor effects, that will be studied in a future separate paper.

\end{document}